\documentclass[
    reprint,        % for double column
    superscriptaddress,
    amsmath,amssymb,
    aps,
    prb,            % change to pra, prl, etc.
]{revtex4-2}

\usepackage{graphicx}   % Include figure files
\usepackage{dcolumn}    % Align table columns on decimal point
\usepackage{bm}         % Bold math
\usepackage{hyperref}   % Hyperlinks
\usepackage{braket}
\hypersetup{
    colorlinks=true,
    linkcolor=blue,
    citecolor=blue,
    urlcolor=blue
}
\begin{document}

\title{Non-Trivial Topological Majorana Architectures: \text{M\"obius} and Trefoil Band Topologies evaluated by Signal to Noise Ratio and Coherence time mesuarements}

\author{Spandan Das}
\email{}
\affiliation{BASIS Independent Upper, Fremont, USA}
\author{Ennis Mawas}
\email{}
\affiliation{Graduate Program in Applied Physics, Northwestern University, Evanston, Illinois 60208, USA}
\date{\today}

% -------------------------------
% Abstract
% -------------------------------
\begin{abstract}
Topological quantum computing is expected to be less sensitive to noise because information is stored in global states rather than local features. \cite{Steiner2020Readout} To see whether different device topologies show measurable differences, we tested three different geometries with separate topological invariants: a Möbius strip, a loop, and a trefoil knot, which have been predicted to work in electronic structures \cite{Venditti2025AngularMZM} \cite{Zheng2022SnSeTopological}. From quantum-capacitance measurements, we extracted power–frequency spectra and fit Lorentzian to obtain linewidth, amplitude, signal-to-noise ratio, and coherence time. Signal to Noise Ratio measures the ratio of the Parity measurement signal to the background noise, an indication of fault-tolerance, and Coherence time measures the time it would take to decohere the quantum state. Across all three topologies, coherence times were similar, with no clear trend linked to the geometry. In contrast, the signal-to-noise ratio was different in the $E_0 = 10 \mu eV, Z=-1$ regime, (Trefoil $>$ \text{M\"obius} $>$ Loop). These observations set a reference point for future experiments designed to more clearly separate topological effects from other device-level parameters.
\end{abstract}

\maketitle

\section{Introduction}
\label{sec:Section 1}
Topological quantum computing aims to build qubits that are less sensitive to noise by using globally protected states \cite{TopologicalQCReview2025}. Majorana zero modes (MZMs) are a leading candidate for this approach \cite{Steiner2020Readout}, but experimental progress has been slow, and results depend strongly on device geometry and fabrication. Nanowire-based topological superconductors provide a tunable platform for Majorana modes \cite{MajoranaNanowire2020}, and recent single-shot interferometric measurements have demonstrated parity readout in InAs–Al devices \cite{Interferometric2025} This raises the question of whether different topological designs can lead to measurable changes in qubit behavior, where previous work has mostly modeled Majorana–quantum dot coupling in basic loop structures and showed that quantum capacitance can reveal parity and other qubit features.

This paper applies the Majorana–Quantum Dot Coupling Model to two nontrivial topologies: a Möbius strip and a trefoil knot. The Hamiltonians are built by imposing geometry-specific boundary conditions, followed by a Schrieffer–Wolff transformation to obtain a parity-resolved form. Quantum capacitance is then calculated as a function of normalized flux to look for qubit-relevant signatures.
A key difficulty is distinguishing real effects from noise. To compare the two topologies, we use two metrics: signal-to-noise ratio (SNR) and coherence time. These are extracted by taking a Fourier transform of the quantum-capacitance data to obtain a power–frequency spectrum and fitting it with a Lorentzian.

The goal of this work is to see whether the $\text{M\"obius}$ strip and trefoil knot offer advantages or behave similarly to simpler structures. This comparison may help identify which geometric features are worth exploring in future device designs. 

Section \ref{sec:Section 2} describes the model and Hamiltonian construction. Section \ref{sec:Section 3} is the analysis of the results achieved. Section \ref{sec:Section 4} discusses the final results and a general analysis. Section \ref{sec:Section 5} discusses similar and previous work.

\section{Formalism}
\label{sec:Section 2}
\subsection{Hamiltonian Construction}
This paper uses the Majorana–Quantum Dot Coupling Model, a Topological Superconductor Hamiltonian with 3 localized modes, $c$, $f$, $d$, and annihilation operators $c^\dagger$, $f^\dagger$, and $d^\dagger$ with assigned hopping amplitudes of $t_{m1}$ and $t_{12}$. The BdG (Bogoliubov-de-Gennes) Hamiltonian of the 1D Nanowire Model is given by:  
\begin{align}
H_0 &= 
2E_M\Big(c^\dagger c-\tfrac12\Big)
+ \Delta_1\Big(f^\dagger f-\tfrac12\Big)
+ \Delta_2\Big(d^\dagger d-\tfrac12\Big),
\\[2pt]
H_{\mathrm{Hop}} &= 
t_{m1}\,f^\dagger(c+c^\dagger)
+ t_{m1}^*(c+c^\dagger)f
+ t_{12}\,f^\dagger d
+ t_{12}^* d^\dagger f .
\end{align}
\par We then performed a Schrieffer-Wolfe Transformation on all $3$ topologies, which can be characterized by the same parity $P$ resolved matrix Hamiltonian with different parameterizations of $E_{M1}$, $t_{L1}$, and $t_{R1}$. \\
The Even Parity Hamiltonian:
\begin{align}
H_{\text{even}} =
\begin{pmatrix}
 -E_{M1}P - \delta E_D 
    & t_{L1} + t_{R1}P \\
 t_{L1}^* + t_{R1}^*P 
    & E_D + E_{M1}P - \delta E_D
\end{pmatrix}
\end{align}
The Odd Parity Hamiltonian:
\begin{align}
H_{\text{odd}} =
\begin{pmatrix}
 E_{M1}P + \delta E_D 
    & t_{L1} - t_{R1}P \\
 t_{L1}^* - t_{R1}^*P 
    & E_D - E_{M1}P + \delta E_D
\end{pmatrix}
\end{align}
The Full Matrix Hamiltonian $4$ x $4$:
\begin{align}
H^{(e/o)} = H_{\text{even}} \oplus H_{\text{odd}} .
\end{align}
\subsection{Phase Dependent Hybridization}
\par Hybridization amplitudes inherit geometric phases associated with both real-space flux and the Bloch bundle, producing an effective Periels phase \cite{Marques2024Peierls} coupled to the Hybridization Energy, $E_m$, and Hopping Amplitudes, $t_{L1, R1}$. The effective Peierls phase contains three contributions: (i) the Aharonov–Bohm phase arising from the extrinsic real-space geometry of the ring, and (ii) the Zak phase, the Berry phase accumulated in the 1D Brillouin zone \cite{MartiSabaté2021ZaksPhase}, which encodes the topological winding of the underlying band \cite{Interferometric2025}, and (iii) A topological phase gained by torsion of each band topology. These phases are as follows:
\begin{align}
\theta_{AB} = \frac{e}{h}\oint{\textbf{A} \cdot\,d\textbf{l}} = 2\pi w \frac{\Phi}{\Phi_0}
\end{align}
where \textbf{A} is the Electromagnetic Gauge Potential, w is the winding number, and $\frac{\Phi}{\Phi_0}$ is the flux throughout the loop.
\begin{align}
\gamma_{\mathrm{Zak}}
= \int_{\mathrm{BZ}} i \langle u_k \mid \partial_k u_k \rangle \, dk
= \frac{1}{2} \int_{\mathrm{BZ}} \partial_k \phi(k)\, dk
= \pi \nu .
\end{align}
where $i \langle u_k \mid \partial_k u_k \rangle$ is the connection of the Berry and $\nu$ is the topological invariant of the topology. 
\begin{align}
\gamma_{\text{topo}}
= \oint A_{\text{geo}}\, ds
= \int_{0}^{L} \tau(s)\, ds. = w\pi
\end{align}
where $A_{geo}$ is the geometric berry connection to encode other topological features of the system like twists in the M\"obius and trefoil geometries, and $\tau(s)$ is the torsion of the geometry. The accumulated geometric phase around the ring is therefore
\begin{align}
\theta= \theta_{AB} + \gamma_{\mathrm{Zak}} + \gamma_{\mathrm{topo}}
\end{align}
We adopt a minimal translationally invariant tight-binding model and, for clarity, set the nearest-neighbor hopping amplitudes uniform throughout the lattice.
\begin{align}
t_{R1, L1} = t_{R, L}\mathrm{exp}(\theta)
\end{align}
Hybridization Hamiltonians, as formulated in \cite{Cheng2009MajoranaSplitting}, consist of Majorana bilinears of the form $i\gamma_i\gamma_j$. Since this operator is Hermitian, the associated hybridization coefficient must be strictly real. As a result, only the real part of the complex tunneling amplitude contributes to the physical splitting:
\begin{align}
E_{ij} = \mathcal{R}(t_{ij})
\end{align}
And using the Phase-dependent formulation of the hopping amplitudes $t_{R1, L1}$ gives
\begin{align}
E_{M1} = \mathcal{R}(t_{R1, L1}) \approx E_{M0} \mathrm{cos}(\theta)
\end{align}
\subsubsection{Closed Loop}
We have now formed concise equations for the Hybridization energy, as well as the hopping amplitudes. Each of these changes is based on the chosen Band Topology of the system. For the Closed Loop, the $w = 0$ and $\nu = 0$, meaning only the AB Phase is present: 
\begin{align}
\theta = \theta_{\mathrm{AB}}
\end{align}
The winding number of the System is $w = 1$, meaning that the total effective Peierls phase is thus:
\begin{align}
\theta_{\mathrm{Loop}} = 2\pi\frac{\Phi}{\Phi_0}
\end{align}
\subsubsection{Möbius Strip}
In addition to the Geometric Phase, a Zak Phase is added due to the number of bands in the topology. Since a Möbius strip involves a flip, there is an amount of torsion proportional to the winding number as well. The effective phase is given by:
\begin{align}
\theta= \theta_{AB} + \gamma_{\mathrm{Zak}} + \gamma_{\mathrm{topo}}
\end{align}
For this system, the winding number, $w = 1$, and the topological invariant is $\nu = 1$, resulting in the effective Peierls Phase:

\begin{align}
\theta_{\mathrm{\text{M\"obius}}} = 4\pi\frac{\Phi}{\Phi_0} + 2\pi
\end{align}

\subsubsection{Trefoil}
The Last Topology chosen for this study is the trefoil knot. This is similar to the Möbius strip in that it involves torsion as well as the Zak and the AB Phase. This is written as:
\begin{align}
\theta= \theta_{AB} + \gamma_{\mathrm{Zak}} + \gamma_{\mathrm{topo}}
\end{align}
But the Trefoil knot has a different  winding number as well as a topological invariant, $w=2$ and $\nu = 2$, making the effective Peierls Phase:
\begin{align}
\theta_{\mathrm{Trefoil}} =  6\pi\frac{\Phi}{\Phi_0} + 4\pi
\end{align}

\subsection{Quantum Capacitance Plots}
The qubit associated with a Majorana zero mode is measured via parity \cite{QuantumDotReadout2016}, which reflects the total number of electrons in the system: odd parity corresponds to $-1$, and even parity to $+1$. The quantum capacitance, which quantifies the change in charge per voltage per available electronic state, can detect changes in this parity \cite{ParityChargeConversion2019} as well as predict Rabi Oscillations \cite{CapParityRabi2024}. Such a change of parity directly corresponds to a qubit transition between $\ket{0}$ and $\ket{1}$, making quantum capacitance a sensitive measure for reading out the state of a Majorana-based qubit. A concise derivation is given to arrive at the Figure 1 plots starting with the General Capacitance Formula,
\begin{align}
C_Q = \frac{Q}{V} = \frac{\partial Q}{\partial V}.
\end{align}

Discretizing this leads to the Total Quantum Capacitance formula given by:
\begin{align}
C_Q^{\rm total} = \sum_{Z = \pm 1} P(Z) C_Q(Z) = \sum \Delta C_Q
\end{align}
where $P(Z)$ is the Boltzmann probability for parity $Z$ \cite{Alterman2017QuantumClassical}.

The probability difference for two parity states at temperature $T$ is:
\begin{align}
P_{\pm} = \frac{e^{-E_{\pm}/k_B T}}{Z}, \quad 
Z = e^{-E_+/k_B T} + e^{-E_-/k_B T}
\end{align}

Expanding the Probability results in the estimate of:
\begin{align}
P_{\pm} \sim \cosh^2\left(\frac{E_{\pm}}{2 k_B T}\right)
\end{align}

And thus the Infintesimal Change of Probability between the 2 Parity States is then:
\begin{align}
\delta P \sim \frac{P_+ P_-}{\left(P_+ + P_-\right)^2} = \frac{\cosh^2\left(\frac{E_+}{2 k_B T}\right) \cosh^2\left(\frac{E_-}{2 k_B T}\right)}
{\left[ \cosh^2\left(\frac{E_+}{2 k_B T}\right) + \cosh^2\left(\frac{E_-}{2 k_B T}\right) \right]^2}
\end{align}

Before finding the $\Delta C_Q$ we have to also find $C_Q(Z=+1)$ and $C_Q(Z = -1)$:
\begin{align}
C_Q(Z) = \frac{\partial Q}{\partial V} 
= e^2 \alpha^2 \frac{\partial f(E_{\rm eff}(Z))}{\partial E_{\rm eff}} \frac{\partial E_{\rm eff}(Z)}{\partial E_D}
\end{align}

\begin{figure*}[h]
    \centering
    \caption{CQ signal as a function of $\phi$ for Closed Loop, $\text{M\"obius}$ Strip, and Trefoil topology}
    \includegraphics[width=0.95\textwidth]{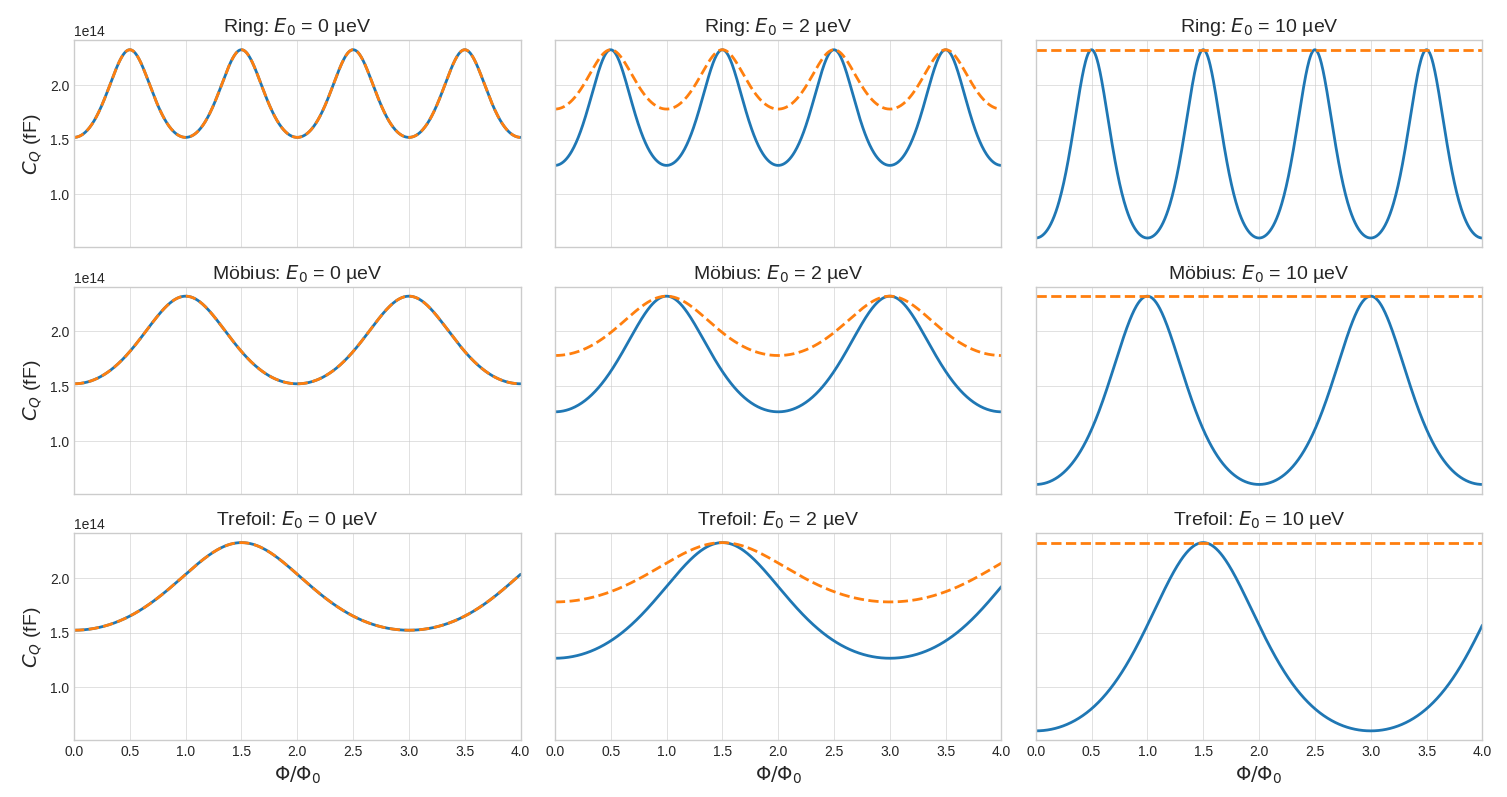}
    \label{fig:CQSignal}
\end{figure*}

Solving this equation using the Parity Resolved Energy Equation, $E_{(Z)} = \sqrt{ (E_0 \pm E_M)^2 + 4 |t_{\rm eff}|^2 }$ results in the Parity Resolved Quantum Capacitance as:
\begin{align}
C_Q(Z) &= e^2 \alpha^2 
\frac{(E_D + Z^2 E_M)^2}{\left[ (E_D + Z^2 E_M)^2 + 4 |t_{\rm eff}|^2 \right]^{3/2}} 
\end{align}
\begin{align}
\times \tanh \left( \frac{\sqrt{ (E_D + Z^2 E_M)^2 + 4 |t_{\rm eff}|^2 }}{2 k_B T} \right).
\end{align}
Putting this all together Results in the Change in Quantum Capacitance from the Parity shift \cite{ParityFlipping2023} as:
\begin{align}
\Delta C_{Q} = - \left[ C_Q(Z=+1) - C_Q(Z=-1) \right] 
\  \delta E_D \frac{4}{k_B T} 
\delta P
\end{align}
This aligns with the Quantum Capacitance found in \cite{Interferometric2025}. \\
Plotting Quantum Capacitance vs. the Flux gives the plots in \ref{fig:CQSignal}.

\subsection{Power vs. Frequency Plots}
\begin{figure*}[h]
    \centering
    \caption{Power as a function of frequency using a Discrete Fourier Transform of $C_Q$ Plot}
    \includegraphics[width=0.95\textwidth]{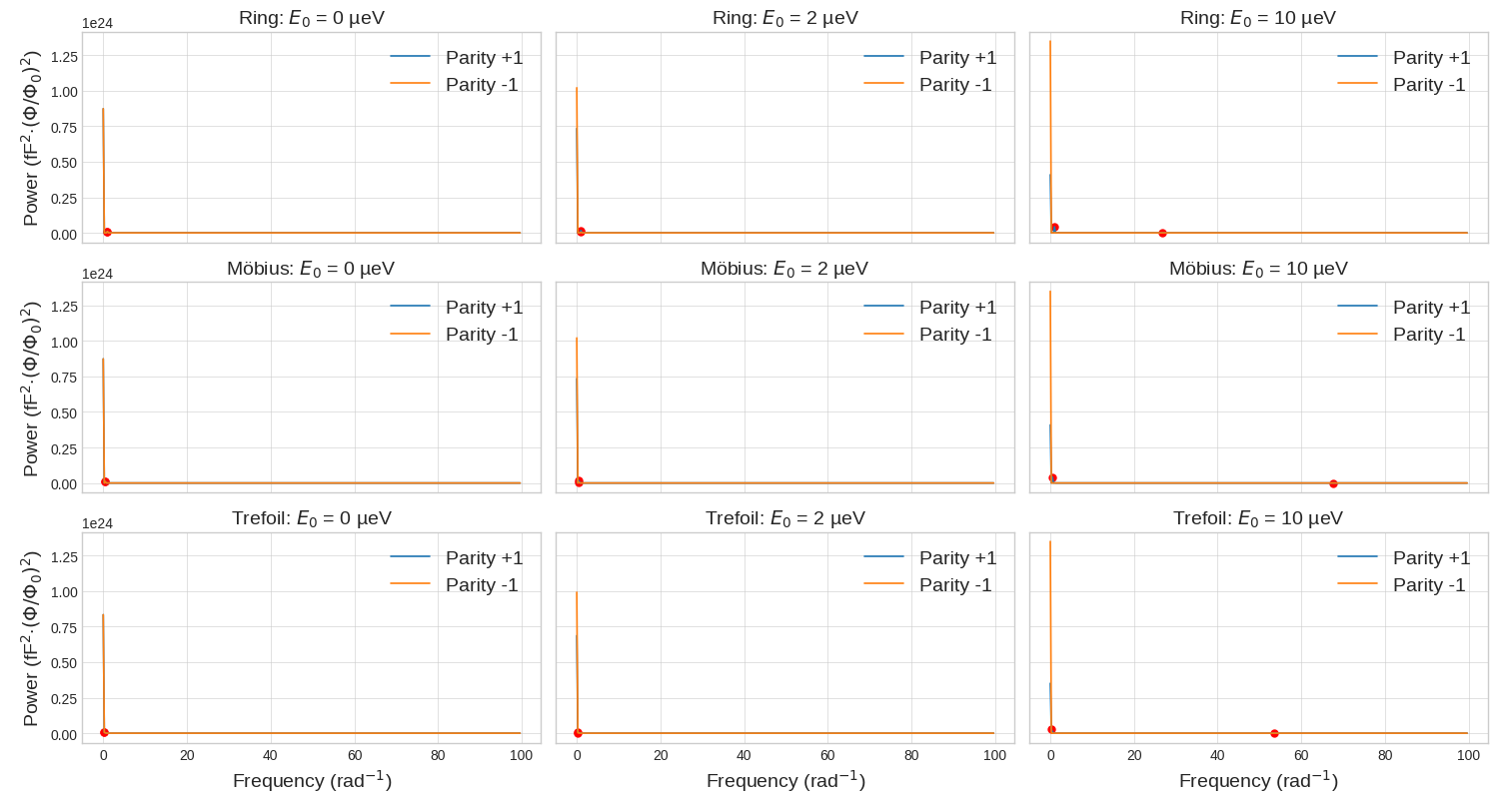}
    \label{fig:PowerSpec}
\end{figure*}

To quantitatively compare the three topologies, we extract the dominant oscillation modes of the quantum capacitance. Because $C_Q$ is naturally periodic in the external flux, we analyze its frequency content by applying a Discrete Fourier Transform (DFT). For numerical stability, we first remove the mean,
\begin{align}
C'_Q[n] = C_Q[n] - \bar{C}_Q,
\end{align}
and apply a Hann window $w[n]$ to suppress spectral leakage,
\begin{align}
C''_Q[n] = w[n]\, C'_Q[n], \qquad \\
w[n] = \tfrac{1}{2} \Big(1 - \cos\frac{2\pi n}{N-1}\Big).
\end{align}
We then compute the DFT
\begin{align}
\tilde{C}_Q[k] = \sum_{n=0}^{N-1} C''_Q[n]\, 
e^{- i 2\pi k n/N}, \\ \qquad k = 0,1,\dots,\frac{N}{2},
\end{align}
and define the (one–sided) power spectral density,
\begin{align}
P[k] = \frac{2}{N^2} \left|\tilde{C}_Q[k]\right|^2.
\end{align}
The peak location,
\begin{align}
\omega_{\mathrm{peak}}
    = \arg\max_{k} P[k],
\end{align}
Plotting the Power vs. Frequency results in \ref{fig:PowerSpec}

\subsection{Signal-to-Noise Ratio Calculation}

The methods of Signal Processing used in this paper are derived from \cite{Sun2022}. For both Signal to Noise Ratio as well as Coherence Time, Calculations can be simplified using a Lorentzian Fit. The general form is given by:
\begin{align}
L(f) = A \, \frac{\gamma^2}{(f - f_0)^2 + \gamma^2} + B
\end{align}
To quantify the visibility of the Lorentzian peak, we define the 
signal--to--noise ratio (SNR) as the ratio between the fitted amplitude 
and the rms noise of the data.  Writing the measured signal as 
\begin{align}
C_Q(f_i) = L(f_i) + \eta_i
\end{align}
where \(\eta_i\) denotes zero--mean Gaussian noise with variance 
\(\sigma^2\), the natural measure of noise strength is the standard 
deviation \(\sigma = \sqrt{\langle \eta_i^2 \rangle}\).  The Lorentzian 
amplitude \(A\) sets the characteristic size of the signal feature, so 
the SNR is
\begin{align}
\mathrm{SNR} = \frac{A}{\sigma}.
\end{align}
This definition matches the standard metric used in spectroscopy, 
resonator readout, and dispersive qubit measurements.
Each Calculation can be found in Table. \ref{tab:lorentzian_results_wide}.

\subsection{Coherence Time Calculation}

Coherence time offers an insight into how long it takes for a quantum computer to be usable, a critical measure of the efficiency of a qubit \cite{Toriyama2025LongT2} \cite{Duttatreya2025PTCoherence}
We derive the Coherence time via. the Lorentzian fit as well. For an exponentially decaying coherence envelope
\begin{align}
s(t) \propto e^{-t/\tau}\,u(t),
\end{align}
its Fourier transform is a Lorentzian whose full width at half maximum (FWHM) is related to \(\tau\) by
\begin{align}
\mathrm{FWHM} = \frac{1}{\pi \tau}.
\end{align}
If the fit parameter \(\gamma\) equals the FWHM (as used in the table), then solving for the coherence time gives
\begin{align}
\tau = \frac{1}{\pi \gamma}. 
\end{align}
The one-sided Fourier transform of \(e^{-t/\tau}u(t)\) is
\begin{align}
\mathcal{F}\{e^{-t/\tau}u(t)\}(f)\propto \frac{1}{1 + (2\pi f \tau)^2},
\end{align}
which is a Lorentzian in \(f\) with FWHM \(=1/(\pi\tau)\). Re-arranging yields \(\tau=1/(\pi\cdot\mathrm{FWHM})\).
Using the fitted \(\gamma\) values from the table:
\begin{align}
\gamma = 0.0997\ \mathrm{Hz} \quad\Rightarrow\quad
\tau = \frac{1}{\pi\cdot 0.0997} \approx 3.19\ \mathrm{s}, \\
\gamma = 0.275\ \mathrm{Hz} \quad\Rightarrow\quad
\tau = \frac{1}{\pi\cdot 0.275} \approx 1.16\ \mathrm{s},
\end{align}
which match the coherence times reported in the table.
This procedure assumes an exponential (single-time-constant) decay of the time-domain signal so that the frequency-domain peak is well described by a Lorentzian. If the decay is non-exponential or dominated by inhomogeneous broadening (Gaussian components), the Lorentzian model and the simple \(\tau=1/(\pi\gamma)\) relation will only give an effective coherence time and should be interpreted accordingly. Each Calculation can be found in Table 1 as well.

\section{Analysis}
\label{sec:Section 3}

\subsection{Quantum Capacitance Plots}

There are a few key features of \ref{fig:CQSignal} that are significant. (i) The independence of Quantum Capacitance in the $E_0 = 10 \mu eV$, $Z=-1$ case for all 3 topologies, (ii) The in distinguishability of the Quantum Capacitance signal in both $Z = -1$ and $Z = 1$ cases for all $E_0 = \mu eV$, and (iii) The proportionality between frequency of Quantum Capacitance with the Topology.

\subsubsection{Quantum Capacitance independence in $E_0 = 10 \mu eV$, $Z=-1$ case}
In the given case, the detuning $E_0$ becomes very large relative to the hopping amplitude $t_{\mathrm{eff}}$ meaning that
\begin{align}
E_{\pm}(\phi) \approx |E_0| 
+ \frac{|t_{\mathrm{eff}}(\phi)|^2}{|E_0|}
, \  |E_0| \gg |t_{\mathrm{eff}}|.
\end{align}
The topology enters the model only through the effective hopping amplitude $t_{\mathrm{eff}}$. However, variations in  $|t_{\mathrm{eff}}|$ produce negligible changes in the ground--state energy $E_{0}$ and therefore in the excitation energies $E_{\pm}$. As a result, the topology does not meaningfully influence the quantum capacitance. Since $t_{\mathrm{eff}}$ is the sole flux-dependent quantity in the Hamiltonian, and its impact on the energy spectrum is minimal, the quantum capacitance remains effectively constant as a function of flux.

\subsubsection{$Z=1, -1$ Symmetry in the $E_0 = \mu eV$ cases}
For this case, the detuning is $E_0 = 0$, which when applied calculate $E_\pm$ gives
\begin{align}
(E_0 + E_M)^2 = (E_0 - E_M)^2, E_0 = 0
\end{align}
For $E_0 = 0\,\mu\mathrm{eV}$, parity—which normally shifts the energy up or down—has no effect on $E_\pm$, making the quantum capacitance parity-independent; consequently, the signatures for $Z = \pm 1$ are indistinguishable, rendering the qubit unreadable in this regime.

\subsubsection{Frequency and Topological Winding number relationship}
We observe that the quantum capacitance $C_Q$ depends solely on the geometric Aharonov--Bohm phase. In this model, its oscillation frequency is $\tfrac{3}{4}\omega$, arising directly from the winding of the geometric phase around the loop. By contrast, the Zak phase and the torsion-induced Berry-connection phase enter the hopping amplitude $t$ only as geometry-dependent constants. Since they do not vary with the external parameter that modulates $C_Q$, they contribute merely an overall phase offset to $t$, leaving its magnitude $|t|$ unchanged. As a result, they do not alter the amplitude or frequency of $C_Q$. Thus, only the geometric Aharonov--Bohm phase produces a parameter-dependent modulation, while the Zak and torsion phases act as constant shifts with no influence on $C_Q$.

\subsection{SNR and Coherence Time Table}

Two features stand out in \ref{tab:lorentzian_results_wide}. (i) All rows except the $E_0 = 10\,\mu\mathrm{eV},\, Z=-1$ case yield identical SNRs, and coherence times, indicating no apparent topology dependence in those regimes. (ii) the coherence time remains fixed at $1.16\,$s for all three topologies in that row, implying that coherence time is insensitive to the underlying band topology. We can also observe that the effectiveness of each topology as a qubit can be ranked, Trefoil > Loop > $\text{M\"obius}$ because of their SNR $(5.26 > 4.89 > 4.31)$\\

\subsubsection{Topology-independence in all except the $E_0 = 10 \mu eV$, $Z=-1$ case}

For all parameter sets except one, the signal-to-noise ratio (SNR) and coherence time $T_2$ are identical across the Trefoil, Loop, and $\text{M\"obius}$ structures. This suggests that either the topological invariants of the underlying band structure are not strongly coupled to either the signal amplitude or noise spectral density.

Formally, if SNR is modeled as
\begin{equation}
\mathrm{SNR} = \frac{|\langle \psi | \hat{O} | \psi \rangle|^2}{S(\omega)},
\end{equation}
where $S(\omega)$ is the noise power spectral density, then topology-dependence requires the expectation value or spectral density to vary with the topological sector $\tau$. The near-constant values across most parameter choices imply
\begin{equation}
\partial_{\tau} \langle \psi | \hat{O} | \psi \rangle \approx 0, \qquad
\partial_{\tau} S(\omega) \approx 0,
\end{equation}
meaning that the observable and its noise environment are topology-agnostic in those regimes.
This is consistent with a phase where the dominant energy scale
\begin{equation}
E \gg \Delta_{\mathrm{top}},
\end{equation}
so that any topological corrections are suppressed.

\subsubsection{Coherence time insensitivity to different topologies}

Despite the SNR variation observed in the above regime, the coherence time remains fixed at $T_2 = 1.16\,\mathrm{s}$ for all three topologies and all parameter values studied. The lack of measurable dependence from the topology on $T_2$ indicates that the dominant contributions to the decoherence rate \cite{Dalton2005Decoherence} $\Gamma_2$,
\begin{equation}
T_2 = \frac{1}{\Gamma_2}, \qquad \Gamma_2 \approx \Gamma_{\mathrm{bath}} + \Gamma_{\mathrm{disorder}},
\end{equation}
are set by environmental or material processes that do not relate strongly to the global topological features.
Formally, this behavior can be expressed as
\begin{equation}
\partial_{\tau} \Gamma_2 \approx 0 \quad \Rightarrow \quad \partial_{\tau} T_2 \approx 0,
\end{equation}
Thus, while topology can affect the quality of the measurement signal under specific conditions, it may not alter the lifetime of coherence, suggesting that decoherence mechanisms are fundamentally insensitive to the band topology in the current system.
\\

\section{Conclusion}
\label{sec:Section 4}

In this work, we have analyzed the quantum capacitance, signal-to-noise ratio (SNR), and coherence time of Majorana-based qubits in three distinct topological geometries: a Closed Loop, a $\text{M\"obius}$ Strip, and a Trefoil Knot. Using parity-resolved effective Hamiltonians derived from geometry-specific boundary conditions, we calculated the flux-dependent quantum capacitance. A Fourier analysis of this signal revealed its dominant spectral components, and Lorentzian fits to the power spectra provided quantitative measures of both the readout SNR and the effective coherence time.

Our findings indicate that there is a clear independence between band topology and intrinsic coherence. Across a large number of parameters, both SNR and coherence time showed little dependence on the underlying band topology, indicating that global topological features do not, by default, dictate these key performance metrics. However, a significant exception emerged under specific conditions, where the SNR became strongly topology-dependent. In this regime, the Trefoil topology yielded the highest SNR, followed by the Loop and $\text{M\"obius}$ geometries. This is caused by the effective Peierls phases, winding numbers, and topological invariants of each geometry, which reshape the frequency distribution of the $C_Q$ signal.

Critically, despite this demonstrable effect on SNR, the coherence time remained remarkably consistent across all topologies and parameter sets. This robust result strongly suggests that decoherence in these systems is not governed by the global topological structure. Therefore, while topological design can be leveraged to enhance the measurability of a qubit state, it may not inherently increase coherence.

Overall, this study establishes that signatures of band topology can indeed be encoded in metrics like SNR under favorable conditions, yet the benefit of extended coherence is not found within this framework. These findings provide a crucial quantitative baseline, reframing the role of geometry in device design: complex topologies should be pursued primarily as tools for improving readout fidelity.

\section{Similar Work}
\label{sec:Section 5}

Previous studies have mostly focused on Majorana–Quantum Dot coupling in simple loop geometries \cite{Interferometric2025}. Several other studies focus on Readout Techniques, an example being Dispersive Readout techniques which have been proposed for high-fidelity Majorana qubit measurements \cite{Dispersive2020}. These works demonstrate how MZMs can be reliably used as qubits. Our study expands on this by considering nontrivial topologies, such as the $\text{M\"obius}$ strip and trefoil knot, which introduce additional geometric phases and winding numbers. By comparing SNR and coherence time across these geometries, we can assess the impact of topological invariants on measurable qubit characteristics.

\begin{table*}[h]
\centering
\caption{Lorentzian fit parameters, SNR, and coherence time $\tau$ for different topologies and bias $E_0$.}
\begin{tabular}{lcccccccc}
\hline
$E_0$ ($\mu$eV) & Topology & Parity & $A$ & $\gamma$ (Hz) & $f_0$ (Hz) & Baseline & SNR & $\tau$ (s) \\
\hline
0   & Loop     & +1 & $7.98\times10^{21}$ & 0.0997 & 2    & $-8.68\times10^{4}$ & 11.9 & 3.19 \\
0   & Loop     & -1 & $7.98\times10^{21}$ & 0.0997 & 2    & $-8.68\times10^{4}$ & 11.9 & 3.19 \\
0   & \text{M\"obius}   & +1 & $7.97\times10^{21}$ & 0.0998 & 3.99 & $-3.42\times10^{4}$ & 11.9 & 3.19 \\
0   & \text{M\"obius}   & -1 & $7.97\times10^{21}$ & 0.0998 & 3.99 & $-3.42\times10^{4}$ & 11.9 & 3.19 \\
0   & Trefoil  & +1 & $7.96\times10^{21}$ & 0.0999 & 5.99 & $-1.70\times10^{4}$ & 11.9 & 3.19 \\
0   & Trefoil  & -1 & $7.96\times10^{21}$ & 0.0999 & 5.99 & $-1.70\times10^{4}$ & 11.9 & 3.19 \\
\hline
2   & Loop     & +1 & $1.38\times10^{22}$ & 0.0997 & 2    & $-8.98\times10^{4}$ & 11.9 & 3.19 \\
2   & Loop     & -1 & $3.69\times10^{21}$ & 0.0997 & 2    & $-8.92\times10^{4}$ & 11.9 & 3.19 \\
2   & \text{M\"obius}   & +1 & $1.38\times10^{22}$ & 0.0998 & 3.99 & $-3.50\times10^{4}$ & 11.9 & 3.19 \\
2   & \text{M\"obius}   & -1 & $3.68\times10^{21}$ & 0.0998 & 3.99 & $-3.34\times10^{4}$ & 11.9 & 3.19 \\
2   & Trefoil  & +1 & $1.37\times10^{22}$ & 0.0999 & 5.99 & $-1.73\times10^{4}$ & 11.9 & 3.19 \\
2   & Trefoil  & -1 & $3.68\times10^{21}$ & 0.0999 & 5.99 & $-1.66\times10^{4}$ & 11.9 & 3.19 \\
\hline
10  & Loop     & +1 & $3.51\times10^{22}$ & 0.0997 & 2    & $-9.98\times10^{4}$ & 11.9 & 3.19 \\
10  & Loop     & -1 & $1.00\times10^{-10}$ & 0.275 & 34   & $5.29\times10^{-13}$ & 4.89 & 1.16 \\
10  & \text{M\"obius}   & +1 & $3.51\times10^{22}$ & 0.0998 & 3.99 & $-3.76\times10^{4}$ & 11.9 & 3.19 \\
10  & \text{M\"obius}   & -1 & $1.00\times10^{-10}$ & 0.275 & 93.9 & $1.82\times10^{-12}$ & 4.31 & 1.16 \\
10  & Trefoil  & +1 & $3.50\times10^{22}$ & 0.0999 & 5.99 & $-1.82\times10^{4}$ & 11.9 & 3.19 \\
10  & Trefoil  & -1 & $1.00\times10^{-10}$ & 0.275 & 63.2 & $1.22\times10^{-12}$ & 5.26 & 1.16 \\
\hline
\end{tabular}
\label{tab:lorentzian_results_wide}
\end{table*}

\bibliographystyle{apsrev4-2}

\end{document}